**Topologically Configurable Nonlinear Vortex Generation at van der Waals Heterostructures**

Hongwei Wang[1], Yuda Wan[2], Kai Wang[2]*, Shuzheng Chen[2], Hao Yan[2], Xu Jiang[1], Xiaodan Lyu[1], Chang-Yin Ji[1], Weibo Gao[1], Peixiang Lu[2, 3*] and Guangwei Hu[1*]

Hongwei Wang and Yuda Wan contributed equally to this work.

Hongwei Wang, Xu Jiang, Xiaodan Lyu, Chang-Yin Ji, Weibo Gao, Guangwei Hu

School of Electrical & Electronic Engineering, Nanyang Technological University, 50 Nanyang Avenue, Singapore

E-mail: guangwei.hu@ntu.edu.sg

Yuda Wan, Kai Wang, Shuzheng Chen, Hao Yan, Peixiang Lu

Wuhan National Laboratory for Optoelectronics and School of Physics, Huazhong University of Science and Technology, Wuhan, China.

E-mail: kale_wong@hust.edu.cn (Kai Wang),lupeixiang@hust.edu.cn (Peixiang Lu)

Peixiang Lu

Hubei Key Laboratory of Optical Information and Pattern Recognition, Wuhan Institute of Technology, Wuhan 430025, China

E-mail: lupeixiang@hust.edu.cn

**Abstract**

van der Waals (vdW) materials offer a highly tunable and efficient platform at nanoscale for nonlinear and quantum optics. Twist-stacked vdW heterostructures enable elegant control of symmetry and interlayer coupling. Prior studies mainly focus on planar twisted interfaces, while neglecting the naturally formed and mandatory defects in such vdW heterostructures. Here, we demonstrate nonlinear singular optics with topologically configurable nonlinear vortex generation at the corner singularity of vdW heterostructures. By tailoring azimuthally discrete second-harmonic phase gradients at each interface, we obtain programmable nonlinear vortex emitters with dominant target OAM components. Nonlinear OAM beams with topological charge $\ell = 1$ and $\ell = -2$ are experimentally realized, respectively. Our work unlocks the untapped potentials of nonlinear singular optics in twisted vdW materials as a reconfigurable and lithography-free platform for nonlinear structured light generation, important in quantum nonlinear optics and related fields.

## 1. Introduction

van der Waals (vdW) materials based on multilayer two-dimensional (2D) crystals offer flexible control of atomic-scale lattice composition and symmetry management[1-12]. The degree of freedoms can be further unleashed via the twisted stacking to form vdW heterostructures, where Moiré effects and interlayer coupling can modify bandstructures and many-body interactions[3,13-22], underpinning various novel phenomena including unconventional superconductivity[23-26], correlated insulating states[27,28] and others.

In nonlinear and quantum optics, vdW heterostructures feature high optical nonlinearity, which is also tunable via twist angle as it regulates crystal symmetry[3,29-35]. For example, the point-group symmetry of vdW heterostructures can be synthesized beyond that of natural crystals, thus rendering new effective nonlinear optical tensors unattainable naturely[20]. Moreover, second harmonic generation (SHG) in multilayer hexagonal boron nitride crystals can be programmed via the twist that modifies superimposition of nonlinear sources at each interfaces[32]. Various unique applications have thus been demonstrated, such as surface enhanced nonlinearity[19,32,36], quantum entangled light sources[3,21,34,37], chiral nonlinear optics[38,39] and programmable SHG engineering[20]. Nonetheless, previous studies mainly focus on nonlinear signal generations remoulded at planar interfaces. By contrast, nonlinear optical responses associated with lateral discontinuities, such as edges, terraces, and corner-like stacking defects, remain comparatively less explored. These geometries provide spatially varying nonlinear phase and amplitude responses that may be exploited for structured nonlinear-light generation.

In this work, we theoretically investigate and experimentally demonstrate nonlinear vortex generation through azimuthally structured vdW stacking geometries (Fig. 1a, b). The materials' thickness and crystalline angle can be tailored, at each section that forms the singularity, to piece-wisely modulate the dynamic and geometric phase of nonlinear signals respectively (Fig. 1c, d). This enables a discrete azimuthal phase-engineering strategy for compact nonlinear vortex emitters with selectable topological charges. Both naturally formed and artificially assembled bilayer and multilayer 3R-$MoS_2$ homogeneous structures are studied (Fig. 1e, f). We achieve the generation of nonlinear vortex beams with topological charges $\ell = -2$ (for 6 layers) and $\ell = 1$ (for 4 layers), respectively. Our method can be extendable to other nonlinear vdW materials and stacking geometries, offering potential routes toward compact structured-light sources for nonlinear optical communications[40-42], sensing[43-45], quantum information processing[21,34,46-49], high-resolution microscopy[50,51] and other applications.

## 2.Results

### 2.1 Principles of singular nonlinear optics near singular defect

We here study the SHG at the defect of vdW 3R-$MoS_2$ heterostructures (see Fig. 1b), which nonetheless can be extended to other nonlinear vdW thin films and other coherent nonlinear processes. Specifically, around these geometric singularities, the structure can be mathematically decomposed as piecewise segment by an azimuthal angular range from $\alpha_{i-1}$ to $\alpha_i$ , where $i = 1,2,\dots,N$ (see Fig. 1c). The inner configuration of each angular pieces (say section $i$) is multilayer planar stacking of nonlinear vdW materials, each layer (for example $j$-th layer, where $j = 1,2,\dots,J_i$) being parametrizable with material choices ($m_{ij}$), thickness ($d_{ij}$) and twisted angle ($\theta_{ij}$), composing a set as $\Pi_{ij} = \{m_{ij}, \theta_{ij}, d_{ij}\}$ with $[\alpha_{i-1}, \alpha_i]$. Here and in this work, all layers are 3R-$MoS_2$ materials, i.e. $m_{ij}$ are all the same, and we neglect this parameter hence. Owing to its non-centrosymmetric $C_{3v}$ symmetry, 3R-$MoS_2$ supports second-order nonlinear optical processes regardless of layer number. Its in-plane crystal lattice is defined by two characteristic directions: the zigzag direction along the *a* axis and the armchair direction along the *b* axis. The second-order nonlinear coefficient of 3R-$MoS_2$ has $\chi^{(2)}_{bbb} = -\chi^{(2)}_{baa} = -\chi^{(2)}_{aab} = -\chi^{(2)}_{aba}$. For consistence and convenience of description, the global *x* coordinate is defined as along zig of first layer ($i$=1, $j$=1), i.e. $\theta_{11} = 0°$.

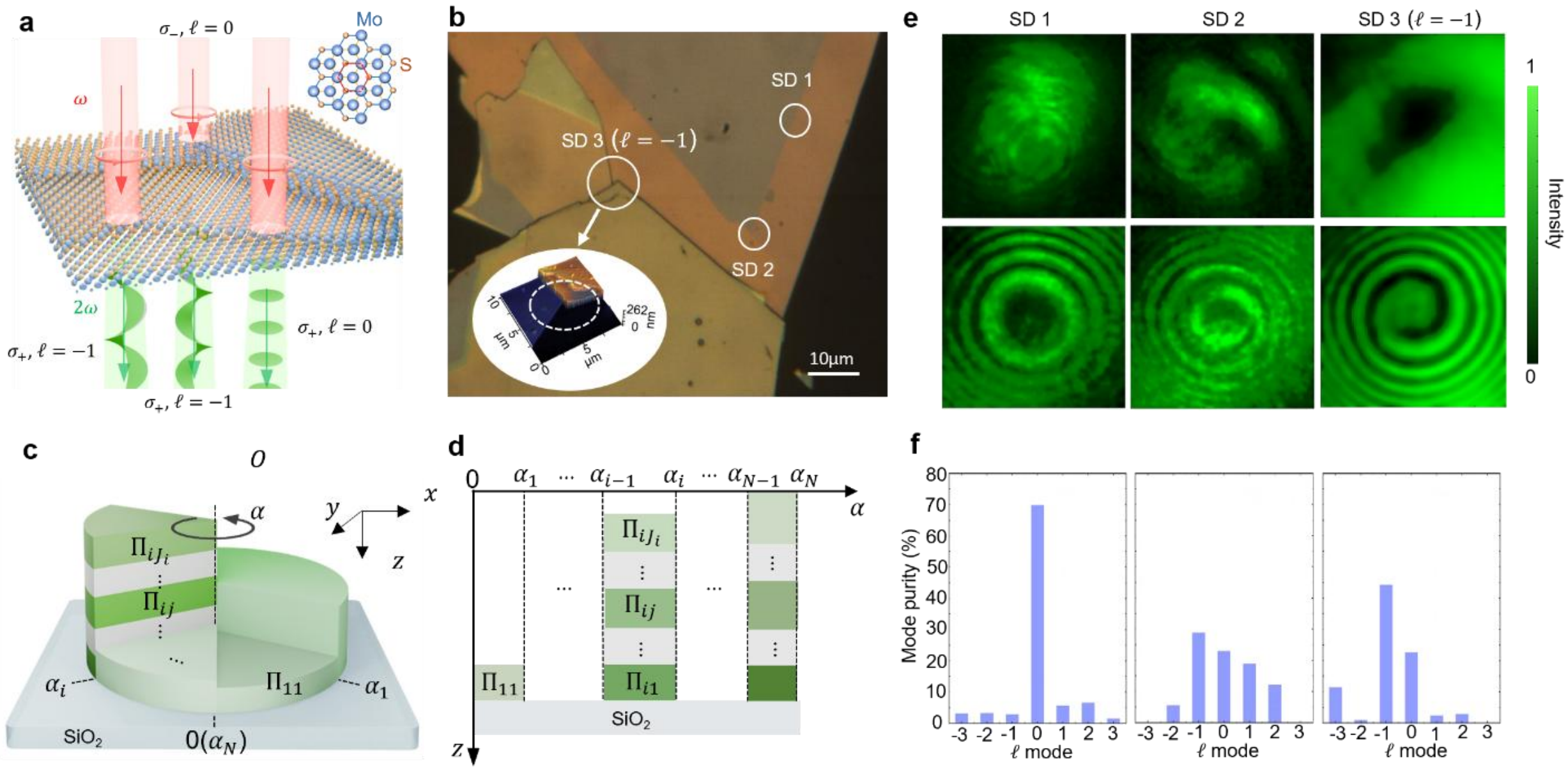

**Figure 1. Concept of nonlinear vortex generation at vdW heterostructural defects. a**, Illustration of the nonlinear vortex generation at corner singularity of 3R-$MoS_2$ homostructures. The inset illustrates the bilayers viewed along the c-axis of the 3R-$MoS_2$. **b**, Top-view optical image of vdW heterostructures and the associated singular defect (SD). SD 1 and 2 indicate defects during mechanical exfoliation. SD 3 marks an artificially stacked nonlinear OAM generator with a topological charge of ℓ = 1. The substrate is $SiO_2$. The inset shows the atomic force microscope (AFM) image near SD 3. The parameters of materials in SD 3, i.e. $\Pi_{11}$ (=$\Pi_{21}$=$\Pi_{31}$), $\Pi_{22}$ and $\Pi_{32}$, are ($d_{11}$, $\theta_{11}$) = (52 nm, 0.22π), ($d_{22}$, $\theta_{22}$) = (57.5 nm, 0.108π) and ($d_{32}$, $\theta_{32}$) = (162 nm, 0.164π), respectively. **c, d** Schematics (c) and angular segmentations (d) of vdWs heterostructures with a corner singularity (O). Here, $m_{ij}$ denotes the material at $j$-th layer of planar structures between the angular range of [$\alpha_{i-1}$, $\alpha_i$]. **e**, The intensity distribution and phase profile of three SDs, respectively. **f**, The experimentally measured mode purity of three SDs, respectively.

To validate our scheme of nonlinear singular optics, we first calculate SHG signals in a planar multilayer, i.e. an angular piece in our configuration (Fig. 1c and Fig. 2a). In a single vdW 3R-$MoS_2$ slab, left-handed circularly polarized light at fundamental frequency ($|\sigma_-, \omega\rangle$) gives rise to right-handed circularly polarized SHG ($|\sigma_+, 2\omega\rangle$). Consider bilayer cases as illustrated in Fig. 2a.

We fix the thickness of bottom layer ($d_1$=55nm) and vary the thickness of second layers ($d_2$) and crystal orientation angle ($\theta_2$). The input is $|\sigma_-, \omega\rangle$. The amplitude and phase distribution of SH signals can be written as follows:

$$E_{2\omega} = \frac{T_2\{(1 - e^{-j\Delta k d_1})e^{j(-\Delta k d_2)} + (1 - e^{-j\Delta k d_2})e^{j(3\Delta\theta)}\}e^{j(k_{2\omega} - 2k_\omega)d_2}}{T_1(1 - e^{-j\Delta k d_1})} = A \cdot \exp(j\phi), \quad (1)$$

where $\Delta k = k_{2\omega} - 2k_\omega$ is the momentum mismatch between the fundamental ($k_\omega$) and second-harmonic ($k_{2\omega}$) waves; $T_1$ and $T_2$ are the amplitude transmission ecoefficiency of the layer 1 and total two layers, respectively. The first and second terms correspond to complex amplitudes of the first and second layers respectively, which exhibit an interference effect (see Supplementary Notes 1 and 2 for more discussions). Note, a relative phase $e^{3\Delta\theta}$ appears, where $\Delta\theta = \theta_2 - \theta_1$ is the angle between the crystalline axis of bilayers, due to nonlinear geometric phase. Besides, a thickness-dependent dynamic phase can be designed according to the momentum mismatch in the quasi-matching condition. Hence, the complex amplitude of nonlinear signals can be engineered over a broad range through the combined control of layer thickness and crystalline orientation.

We first plot the normalized SHG intensity $A$ in Fig. 2b, defined as the ratio of the SHG intensity generated by the bilayer structure to that generated by the first layer alone. The normalized SHG intensity exhibits a peak at ($d_2$=94 nm, $\Delta\theta$=102°) and a valley at ($d_2$=130 nm, $\Delta\theta$=50°), arising from constructive and destructive interference, respectively, between the SHG signals generated in the two layers. Interestingly, two interference conditions happen in $\Delta\theta$-dimension with a spacing close to 60°, due to the emerging nonlinear geometric phase as $e^{3\Delta\theta} \approx -1$. A slight difference from 60° is due to dynamic phases as the result of momentum mismatch[52]. We further calculate the transmission phase ($\phi$) as illustrated in Fig. 2c. Importantly, we can obtain second-harmonic iso-amplitude contour lines in Fig. 2b, along which $\phi$ can be swept from 0 to 2π (the inset of Fig. 2c).

This result validates a key design principle for nonlinear singular optics in vdW heterostructures: phase can be independently controlled by selecting stacking parameters while maintaining similar SH output amplitudes. Note such iso-amplitude full-2π phase manipulation can be extended to more layers (see Supplementary Note 2). This suggests the programmable nonlinear wavefront shaping in each angular piece around the singular defect (SD), underpinning the foundation of singular nonlinear optics at vdW heterostructures.

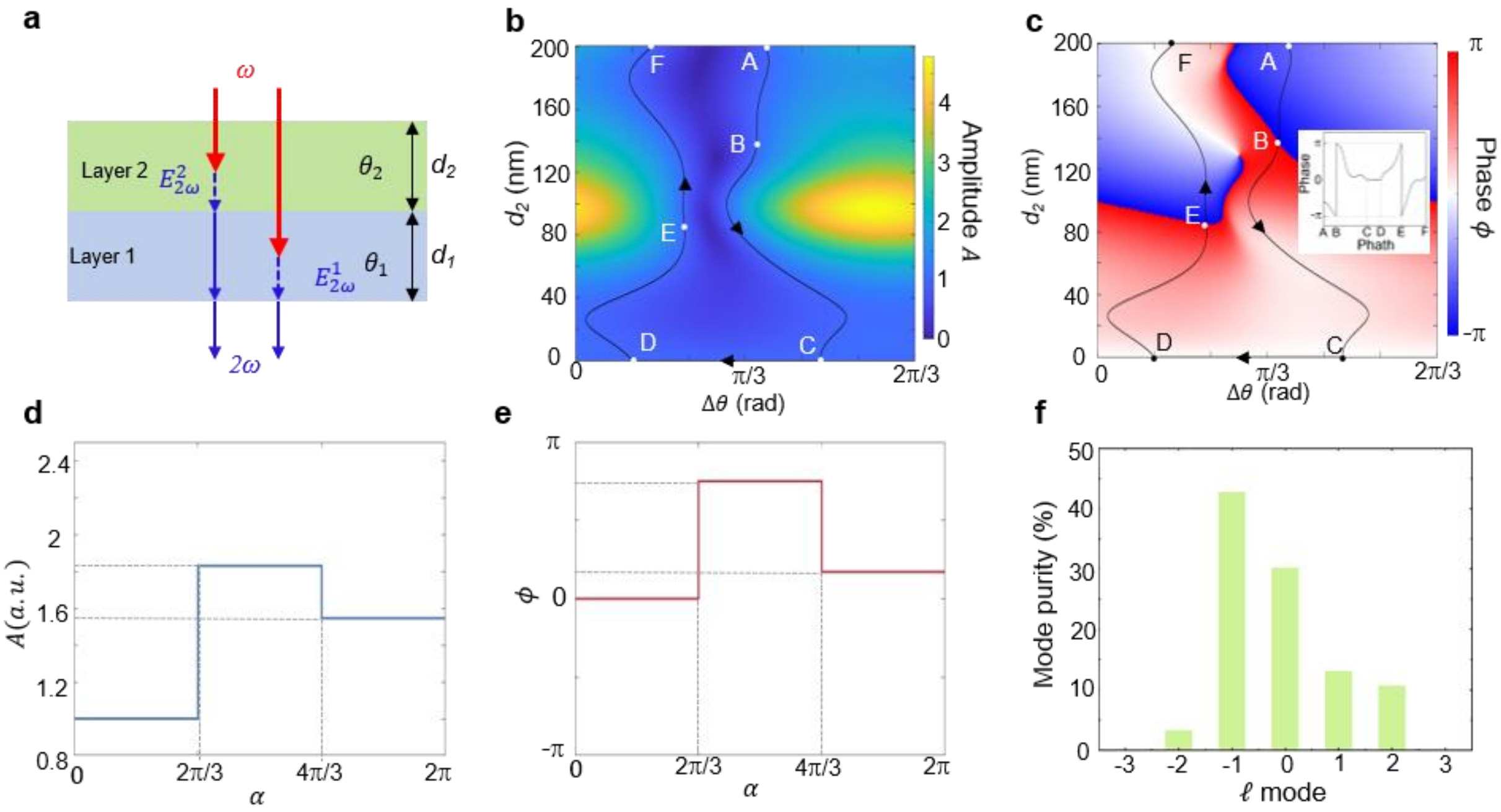

**Figure 2. Principles of nonlinear vortex generation based on 3R-MoS₂ stacks. a**, The schematic of a twisted bilayer 3R-MoS₂ stack. The purple dashed line denotes the second-harmonic signal generated in the current layer, whereas the solid line represents the transmitted second-harmonic signal. **b**,**c** The amplitude ($A$) and phase ($\phi$) distributions of the SHG are calculated for a two-layer rotationally stacked structure. With the bottom layer thickness fixed at $d_1$ = 52 nm, the second layer thickness ($d_2$) and crystal orientation angle ($\theta_2$) are varied to investigate their effects on the SHG response. The inset (**c**) shows the trend of phase variation along the contour lines. **d,e** The calculated $A$ and $\phi$ of the SD 3 along the azimuthal angular $\alpha$. **f**, The calculated mode purity of the SD 3.

To experimentally demonstrate our scheme, we further analyze three representative SD in Fig. 1 (see methods for sample preparation). At SD 1, two discrete second-harmonic phase segments are formed as $(\phi_1, \phi_2)$ = (0, 0.2π). Such phase contrast partially disrupts the azimuth symmetry and produces a weak orbital angular momentum (OAM) generation. However, due to small phase contrast, the emission field remains primarily circularly polarized with dominating zero OAM (Fig. 1f). At SD 2, three discrete height differences introduce an azimuthal phase segmentation capable of generating a nonlinear vortex, but the lack of control over phase and transmittance in randomly exfoliated samples leads to low mode purity. In contrast, the artificially stacked multilayer at SD 3 provides deterministic phase control (see AFM image in the inset, Fig. 1b). For SD 3, the calculated discrete transmission phase $(\phi_1, \phi_2, \phi_3)$ = (0, 0.75π, 0.172π), suggesting a modal purity (up to 43%) with $\ell = -1$ as shown in Fig. 2 d-f. Theoretically, at least three independent phase pixels, e.g. $\phi$ = (0, 2π/3, −2π/3) are required to generate an $\ell$ = 1 vortex with modal purity exceeding 70% (Supplementary Notes 3, 4). Accordingly, the SHG emission at N = 2 carries zero OAM, whereas increasing the number of phase segments to N = 3 produces a clear $\ell$ = 1 vortex, as confirmed by the mode profile, phase distribution, and interference pattern in Fig. 1e.

## 2.2 Experimental nonlinear vortex beam generation

To enhance mode purity, we introduce a simplified four-step stacking for ℓ = 1 OAM generation, with a mode purity up to 84% theoretically (Fig. 3a). Here, this configuration is constructed through a step-by-step transfer process in which rectangular flakes are sequentially stacked to form the target four-layer geometry. Starting from the first layer as the base, the second layer is transferred onto it, followed by the third layer stacked with a designed orientation, and finally the fourth layer is placed to complete the structure. Throughout this process, the orientation of each rectangular flake is defined in the same in-plane coordinate system, with all crystallographic phase angles referenced to a common global $x$-direction. Each layer is therefore characterized by an absolute angle $\theta_i$ relative to this $x$-axis, rather than by a local or pairwise relative angle, which ensures an unambiguous angular definition and a controllable nonlinear phase distribution in the final device. The optical image and AFM image of the fabricated nonlinear optical vortices generator have been shown in Fig. 3b. Such method thus relaxes fabrication constraints and requires less angular precision, as illustrated in Fig. S4 (see Supplementary Note 3).

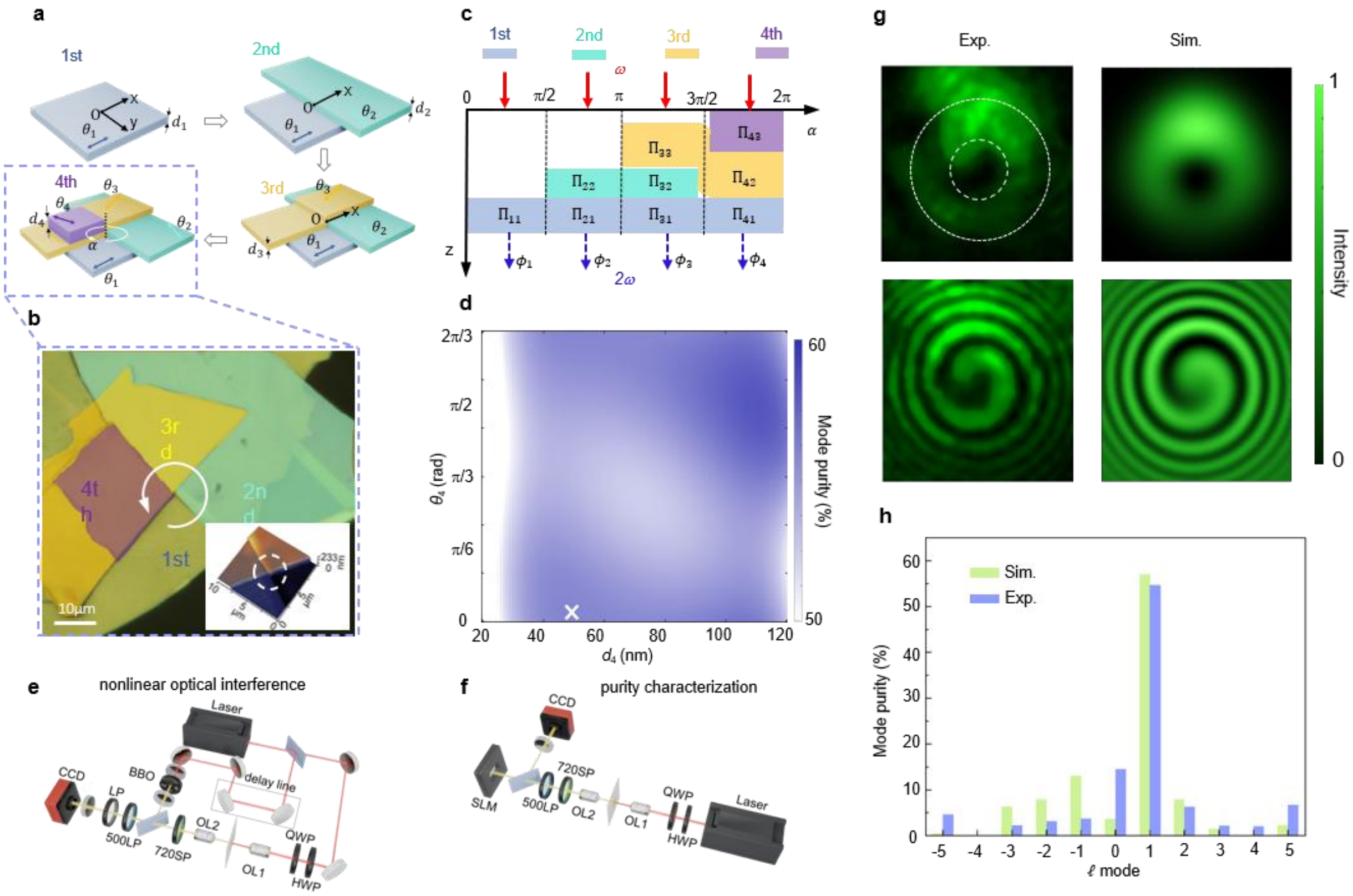

**Figure 3. Optimization of optical vortices with ℓ = 1. a**, Schematic illustration of the sequential transfer process for the four-step stacked nonlinear vortex generator. **b**, The optical image and AFM image of the fabricated nonlinear optical vortices generator. **c**, Cross-sectional schematics of optical elements. Here, the structural parameters satisfy $\Pi_{11} = \Pi_{21} = \Pi_{31} = \Pi_{41} = (d_1, \theta_1)$, $\Pi_{22} = \Pi_{32} = (d_2, \theta_2)$, $\Pi_{33} = \Pi_{42} = (d_3, \theta_3)$ and $\Pi_{43} = (d_4, \theta_4)$. **d**, Calculated modal purity of the second-harmonic OAM state as a function of crystalline phase angle $(\theta_4)$ and thickness of the layers 4 $(\theta_3)$ and 4 respectively. The thickness and crystal phase angle distribution are fixed as $(d_1, d_2, d_3,)$ = (84 nm, 60 nm, 64 nm) and $(\theta_1, \theta_2, \theta_3)$ = (-0.167π, 0.25π, 0.056π). The white cross indicates the parameters selected in our experiment. **e**,**f**, Experimental setup for nonlinear optical interference and purity characterization. HWP: half-wave plate; QWP: quarter-wave plate; OL: objective lens; 500LP: long-pass filter at 500nm; 720SP: short-pass filter at 720nm; LP: linear

polarizer. **g**, The measured and calculated intensity profile and interference pattern of the four-step-phase nonlinear optical vortices generator with ℓ = 1. **h**, The experimental and simulated purity of the SH beam intensity for OAM state with ℓ = 1.

Figure 3c presents a cross-sectional schematic of the four-step stacked nonlinear vortex generator, for which the crystallographic phase angles satisfy $\theta_{11} = \theta_{21} = \theta_{31} = \theta_{41} = \theta_1$, $\theta_{22} = \theta_{32} = \theta_2$, $\theta_{33} = \theta_{42} = \theta_3$ and $\theta_{43} = \theta_4$. By optimizing the stacking parameters, the nonlinear phase contribution is designed to approximate the discrete azimuth phase points $(\phi_1, \phi_2, \phi_3, \phi_4) = (0, \pi/2, \pi, -\pi/2)$. For example, in the experiment, fixing the crystallographic phase angles of the first three layers to $(\theta_1, \theta_2, \theta_3) = (-0.167\pi, 0.25\pi, 0.056\pi)$ gives discrete nonlinear phase points $(\phi_1, \phi_2, \phi_3) = (0, 0.55\pi, 1.33\pi)$. Then, the parameters of the fourth layer were adjusted to balance the relative amplitude and phase of the nonlinear emission, thereby achieving constructive interference in the target mode and suppressing unwanted components. The optimized parameters marked in Fig. 3d yield a target mode purity above 57%, corresponding to $(d_4, \theta_4)$ = (50 nm, -0.16π ); see comprehensive optical characterization in Methods and Supplementary Note 5.

To characterize the nonlinear vortex beam generation, a Mach-Zehnder interferometer at free space is constructed between emitted SH signals and reference beam generated from a BBO under collinear conditions (Fig 3e). The interference pattern is collected by an EMCCD. For purity characterization shown in Fig. 3f, the signal beam is modulated by a spatial light modulator encoded with matching filters of basic Laguerre-Gaussian modes with different topological charge ($LG_{0,l}$). The on-axis intensity of the output beam shows the correlation factor between the signal field and those basic modes, which can be used as purity information (see Supplementary Note 5).

The device is illuminated by a circularly polarized 1240 nm laser beam with a 5 µm spot radius, focused onto the region indicated by the white circle in Fig. 3b. We record donut-shaped intensity profile with a clear central null (top left, Fig. 3g). Interferometric measurements reveal a single-arm spiral fringes (bottom left, Fig. 3g), confirming a topological charge of ℓ = 1 and agree well with simulations (right panels, Fig. 3g). Furthermore, the measured mode purity for ℓ = 1 exceeds 54.7 %, with residual contributions of 6.3 % (ℓ = 2), 14.2 % (ℓ = 0), and 6.67 % (ℓ = 4). These agree with theoretical predictions, confirming that configurable nonlinear OAM beams can be generated. Compared with the three-lobe stack in Fig. 1, the four-lobe strategy yields significantly higher mode purity, which is essential for high-fidelity nonlinear OAM generation.

### 2.3 Topologically tunable nonlinear vortex generation

Our approach can allow the customized and tunable nonlinear vortex generation, and, as a proof of concept, we present a nonlinear vortex with a topological charge of ℓ = −2. Theoretical analysis reveals that when the number of discrete phase points exceeds six, a maximum mode purity of ℓ = −2 up to 73% can be achieved under ideal amplitude and phase control conditions (see Supplementary Note 3 and 4). Based on this insight, we design a six-segmentation architecture (Fig. 4a). Specifically, the 2nd and 6th layers are assembled with an angle of 60°, while 3rd, 4th, and 5th layers are selected after exfoliation to exhibit an opening angle of 120°, forming a quasi-hexagonal phase structure via rotational stacking (see methods for fabrication details).

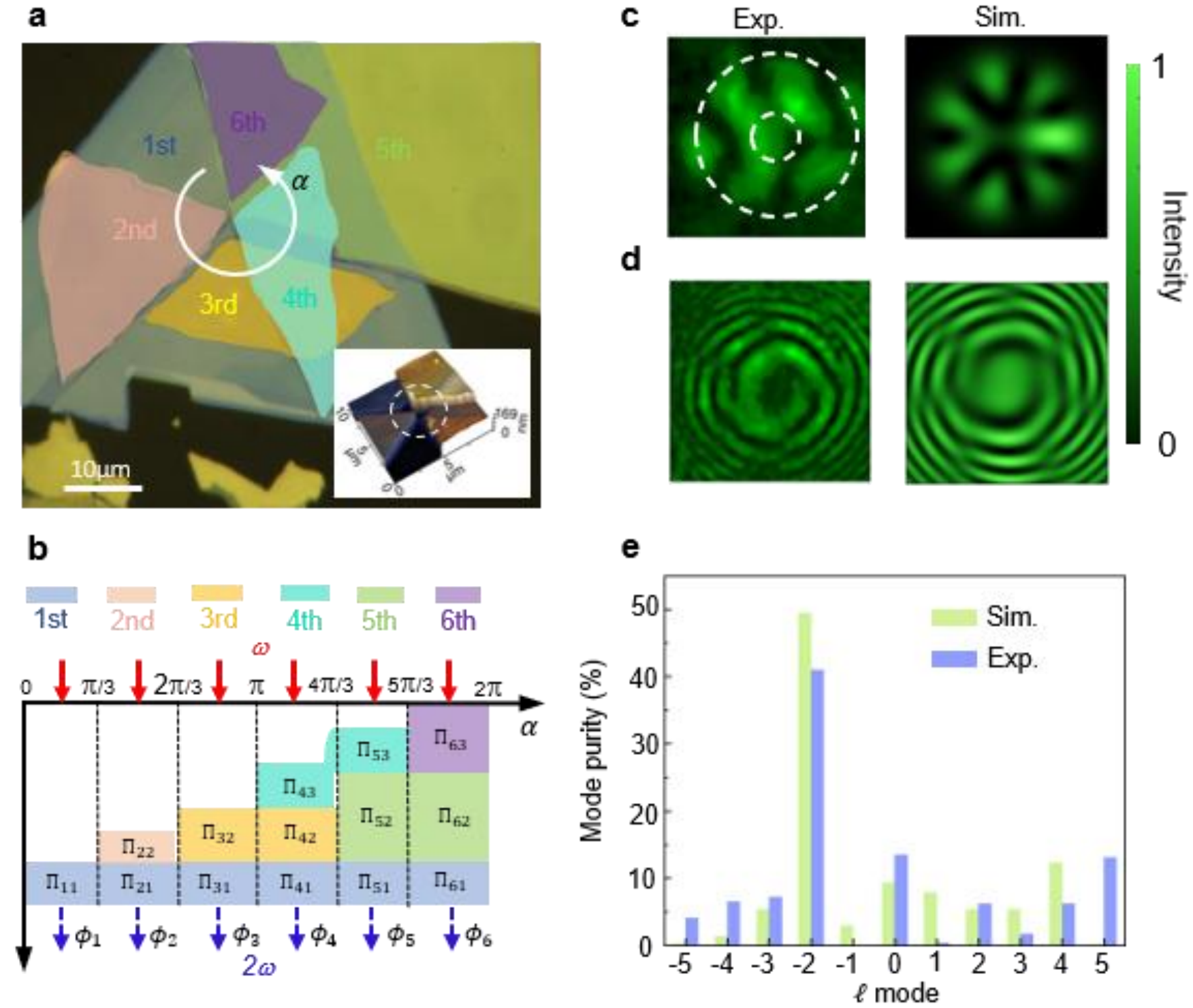

**Figure 4. Generation of optical vortices with ℓ = −2. a**, The optical image and AFM image of the fabricated nonlinear optical vortices generator with ℓ = −2. The $\theta_1 - \theta_6$ are the crystalline phase angles of the layers 1 - 6, respectively. **b**, Cross-sectional schematics of the 3R-$MoS_2$ nonlinear optical vortex generator with asymptotic angle of the phase gradient $\Delta\alpha = \pi/3$ . Here, the structural parameters satisfy $\Pi_{11} = \Pi_{21} = \Pi_{31} = \Pi_{41} = \Pi_{51} = \Pi_{61} = (d_1, \theta_1)$, $\Pi_{22} = (d_2, \theta_2)$, $\Pi_{32} = \Pi_{42} = (d_3, \theta_3)$, $\Pi_{43} = \Pi_{53} = (d_4, \theta_4)$, $\Pi_{52} = \Pi_{62} = (d_5, \theta_5)$ and $\Pi_{63} = (d_6, \theta_6)$. Here, the optimized geometric parameters are: $(d_1, d_2, d_3, d_4, d_5, d_6)$ = (77 nm, 77 nm, 51 nm, 65 nm, 37 nm, 21 nm) and $(\theta_1, \theta_2, \theta_3, \theta_4, \theta_5, \theta_6)$ = (0, 0.056π, -0.017π, 0.028π, 0.028π, 0.028π). **c, d** The experimentally obtained spatial profile (left; panel c) and the interference pattern (left, panel d), compared with the corresponding simulation results (right, panel c and d) for the six-step-phase OAM state with ℓ = −2. **e**, The experimentally recorded purity of the donut in c.

In this configuration, SDs of the second and third assembled layers overlap along the vertical axis, giving rise to a vertically distributed topological core, as shown in Fig. 4b. Using a multilayer nonlinear coupling model, we theoretically determine the optimal thickness and crystal orientation angle for each layer to achieve the highest mode purity (see Supplementary Note 4). This configuration yields a calculated OAM mode purity of 49.45% for ℓ = −2. Experimentally, a petal-like pattern is recorded (top left, Fig. 4c), with excellent agreement with an ideal ℓ = −2 vortex beam (top right). Besides, with the interference via a planar reference wave (bottom left), a double-helix structure near the centre is obtained, further validating a topological charge ℓ = −2 (bottom right).

Figure 4d shows the experimentally measured mode purity of the device. The coefficient for the ℓ = −2 mode is approximately 42%, with residual contributions primarily from ℓ = 0, 5, 4, −3, and

−4. The minor residual mainly originates from the non-uniform and hence non-ideal phase gradient caused by stacking defects during material selection and manufacturing. The measured mode distributions show good agreement with the simulation, validating the effectiveness of nonlinear singular optics by adjusting the folding angle, thickness, and crystal orientation of stacked vdW materials for efficient and most importantly, topologically tunable nonlinear vortex generation.

## 3. Discussion

In summary, we present the nonlinear singular optics that naturally exists and is also artificially manipulable in vdW heterostructures for topologically tunable nonlinear vortices generations. This structural modulation enables the creation of discrete angular SHG phase gradients, thereby forming programmable orbital angular momentum and structured light sources. Experimentally, we achieved OAM beams with topological charges $\ell = 1$ and $\ell = -2$, with measured mode purities reaching 54.7 % and 42 %, respectively.

Compared with previously reported nonlinear vortex beam generation schemes based on OAM-pumped nonlinear conversion[53-55], ferroelectric topological textures for nonlinear vortex generation[56], intrinsic spin-orbit conversion[57], geometric-phase nanostructuring[58-60], our approach directly engineers the phase and amplitude of SH signals through twist-controlled stacking, enabling vortex generation under simple Gaussian excitation without bulky optical systems, stringent material-symmetry requirements, or complex nanofabrication. This intrinsic and lithography-free strategy is broadly applicable to common vdW materials and provides a scalable platform for higher-order OAM states, improved mode purity, and extension to other nonlinear and quantum light-generation processes. Extending this concept to higher-order stacks, resonant photonic structures, and quantum nonlinear processes promises to build compact integrated platforms for advanced classical and quantum photonic applications.

## 4. Materials and methods

**Sample fabrication.** The 3R-$MoS_2$ flakes were mechanically exfoliated onto polydimethylsiloxane (PDMS) gel film and then dry transferred onto 300-μm-thick quartz substrates with a custom-built motorized transfer stage. To fabricate the nonlinear vortex generator, a large 3R-$MoS_2$ flake was first dry transferred onto 300-μm-thick quartz substrates with a custom-built motorized transfer stage. Then smaller flakes were transferred onto the large flake side by side with the help of the alignment function of the CCD camera on the transfer stage. After transferring, the AFM image was obtained by VistaScope microscope from Molecular Vista Inc combined with SurfaceWork 3.0.

**The optical characterization.** The SHG interference was performed with a custom-built Mach-Zehnder interferometer. The sample was excited by a pump beam at 1240 nm from a tunable femtosecond source. The SHG signal and the frequency doubled reference beam were collected into a charge coupled device to obtain the interference profile. For purity measurement, the SHG signal was directed to a spatial light modulator to analyse its overlap with the pure Laguerre-Gaussian modes. See more details in Supplementary Note 5.

**Acknowledgements.**

The work is supported by National Research Foundation of Singapore under award no. NRF-CRP31-0001. G. Hu acknowledges the Nanyang Assistant Professorship Start-up Grant, Ministry of Education (Singapore) under AcRF TIER2 (MOE-T2EP50224-0017 and MOE- T2EP50125-0013), and A*STAR under its MTC IRG Grant (Project No. M24N7c0087). H. W. acknowledges research funding by the National Natural Science Foundation of China (NSFC) (62305210). K. W. acknowledges National Natural Science Foundation of China (No. 12274157, No. 12021004, No. 12274334, No. 11904271). Special thanks are given to the Analytical and Testing Center of HUST, the Experiment Center for Advanced Manufacturing and Technology in the School of Mechanical Science & Engineering of HUST, and the Center of Micro−Fabrication and Characterization (CMFC) of WNLO for use of their facilities.

**Author contributions**

H. W. K. W. and G. H. conceived the idea. Y. W., H. Y., S. C. and K. W. performed the experiment on fabrication and optical measurement. H. W., X. L., W. G., X. J. and G. H. performed the theoretical analysis. H. W., C. J. and G. H. performed the numerical calculation and simulations. G. H., K. W. and P. L. supervised the project. All authors contributed extensively to the interpretation of the results and to the writing of the manuscript.

**Data availability:**

Data supporting the findings of this study are available upon request from the corresponding author. The original data are provided with this paper.

**Conflicts of interest**

The authors declare no competing interests.